\documentclass[3p,times,twocolumn]{elsarticle}

\usepackage{ecrc}

\volume{00}

\firstpage{1}

\journalname{Nuclear Physics B Proceedings Supplement}

\runauth{Markus Wobisch}

\jid{nuphbp}

\jnltitlelogo{Nuclear Physics B Proceedings Supplement}

\usepackage{amssymb}

\usepackage[figuresright]{rotating}

\newcommand{\Rtt}{R_{3/2}}

\newcommand{\ptmax}{p_T^{\rm max}}
\newcommand{\ptmin}{p_T^{\rm min}}

\newcommand{\chijj}{\chi_{\rm dijet}}

\newcommand{\as}{\alpha_s}

\newcommand{\asmz}{\alpha_s(M_Z)}
\newcommand{\asmur}{\alpha_s(\mu_R)}
\newcommand{\aspT}{\alpha_s(p_T)}

\newcommand{\pythia}{{\sc pythia}}

\newcommand{\fastnlo}{{\sc fastnlo}}
\newcommand{\nlojet}{{\sc nlojet++}}
\newcommand{\Rcone}{R_{\rm cone}}

\newcommand{\ppbar}{p{\bar{p}}}

\def\met{{\mbox{$E\kern-0.42em\raise0.15ex\hbox{/}_{T}$}}}

\begin{document}

\begin{frontmatter}



\dochead{}

\title{Recent QCD results from the Tevatron}
\author{Markus Wobisch}
\address{Louisiana Tech University, Ruston, LA, USA}

\begin{abstract}
Recent QCD related results from the CDF and the D\O\ experiments
are presented based on proton anti-proton collision data
at $\sqrt{s}=1.96\,$TeV, taken in Run II of the Fermilab Tevatron Collider.
Measured observables include inclusive photon and diphoton production,
vector boson plus jets production, event shape variables, and inclusive multijet production.
The measurement results are compared to QCD theory calculations in 
different approximations.
A determination of the strong coupling constant from jet data
is presented.
\end{abstract}


\end{frontmatter}


\section{Introduction}
\label{sec:intro}

This presentation was given two days before 
the Fermilab Tevatron Collider had stopped its operations.
During the previous ten years, in Run II, the Tevatron
had delivered an integrated luminosity of 12\,fb$^{-1}$
of proton-antiproton collisions at a center of mass energy
of 1.96\,TeV to the CDF and D\O\ experiments,
with peak luminosities of up to 
$4.3 \cdot 10^{32}\,{\rm cm}^{-2}{\rm s}^{-1}$.

This article presents an overview of recent QCD results from 
the CDF and the D\O\ experiments, based on data sets
with integrated luminosities of $0.7$ -- $8.2$\,fb$^{-1}$. 
Presented are measurements of inclusive isolated photon and
diphoton production,
vector boson plus jet production for different jet multiplicities,
and event shape variables.
Also presented are measurement results of multijet production
together with phenomenological analyses to
constrain the parton distribution functions (PDFs) of the proton
and to determine the strong coupling constant, $\as$.
In all cases, the data are corrected for instrumental effects and are 
presented at the ``particle level,'' which includes all stable particles 
as defined in Ref.~\cite{Buttar:2008jx}.
The results are used to test either particle-level predictions
by Monte Carlo events generators, or perturbative QCD (pQCD) 
calculations in fixed order in $\as$ which are corrected for 
non-perturbative effects.

\section{Photon Production}
\label{sec:photon}

Photon cross sections in hadron collisions receive contributions
from ``prompt'' photons which directly emerge from the hard subprocess, 
and from photons which are produced in the fragmentation 
of energetic $\pi^0$ and $\eta$ mesons.
The latter are usually accompanied by hadrons, and their contribution 
can be significantly reduced by requiring the photon to be isolated 
from other particles in the event.
Isolated photon cross sections are therefore dominated by prompt photons.
At lowest order, prompt photons are produced via quark-gluon Compton 
scattering or quark-antiquark annihilation,
and are therefore directly sensitive to the dynamics of the hard subprocess
 and to $\as$ and the PDFs of the hadrons.
Furthermore, diphoton final states are also signatures for various 
new physics processes, such as extra spatial dimensions and for 
heavy new particles, such as the Higgs boson, decaying into photons.

\subsection{Inclusive Isolated Photon Production}

Photon production has been considered an ideal source of direct 
information on the gluon density in 
the proton~\cite{Lai:1996mg,Martin:1996ev}.
However, it was observed~\cite{Aurenche:1998gv} that not all 
experimental data are consistently described by pQCD
calculations in fixed order of $\alpha_s$.
On the one hand, it was argued that the existing data may not 
be consistent~\cite{Aurenche:1998gv}.
It was also suggested that the phenomenological introduction of 
an intrinsic transverse momentum of the incoming partons may help 
improve the description of the data and reduce the 
inconsistencies~\cite{Martin:1998sq}.
But this ad-hoc procedure still had a significant model dependence
and was not seen to be a fundamental solution.
In recent global PDF fits photon data have been 
excluded~\cite{Martin:2009iq,Lai:2010vv,Ball:2011mu}.

In order to rescue the photon data as a source of additional 
information on the PDFs, it is important either to identify
critical missing pieces in theory or to clearly establish 
inconsistencies of existing data sets.
Precision measurements in new kinematic regions are vital for 
testing theory predictions.

\begin{figure}
\centering%
\includegraphics[scale=0.35]{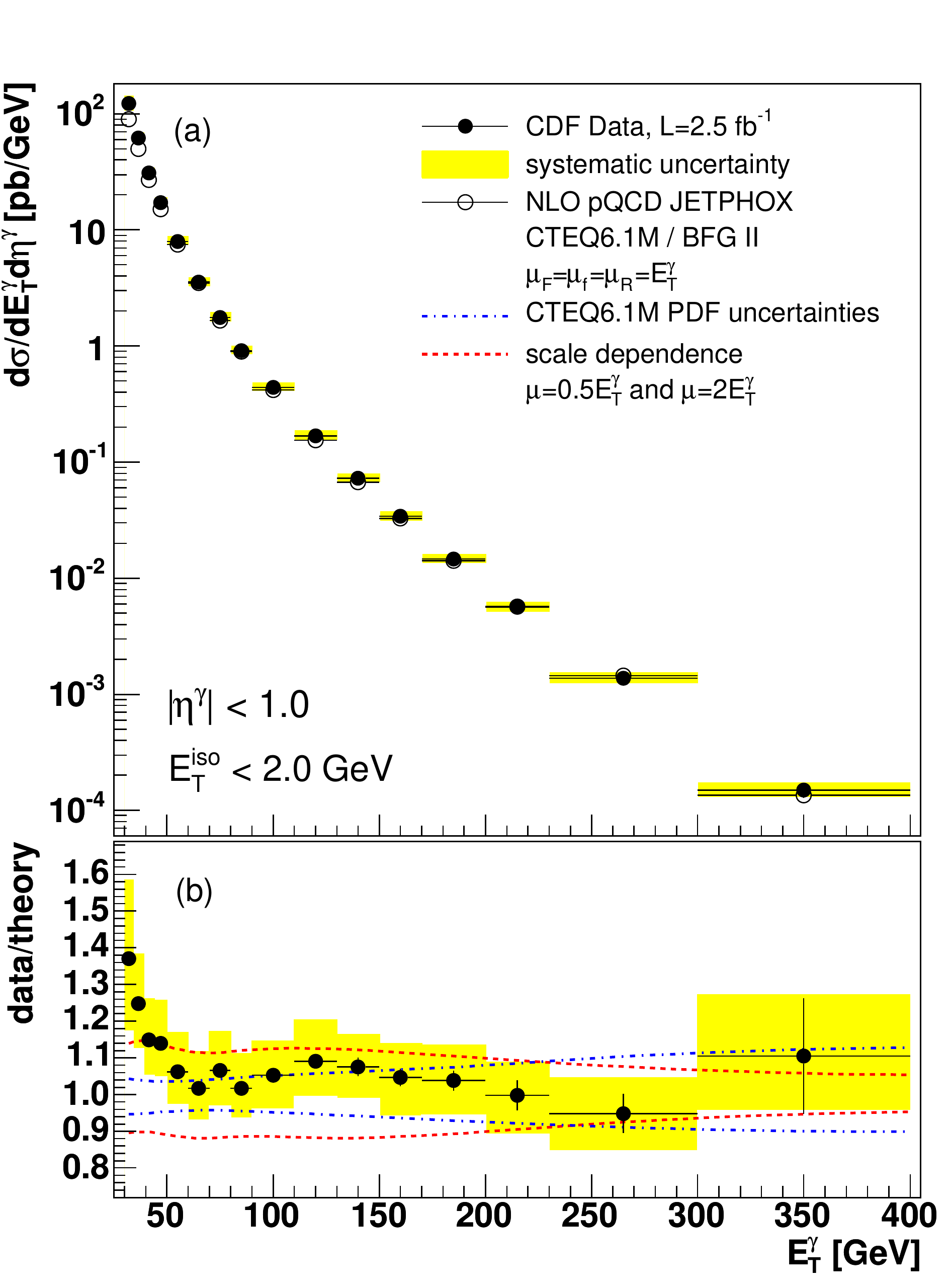}
\caption{CDF measurement of the inclusive isolated photon cross section,
measured differentially in photon $E_T$.
\label{fig:cdfinclphoton1}}
\end{figure}

\begin{figure}
\centering%
\includegraphics[scale=0.4]{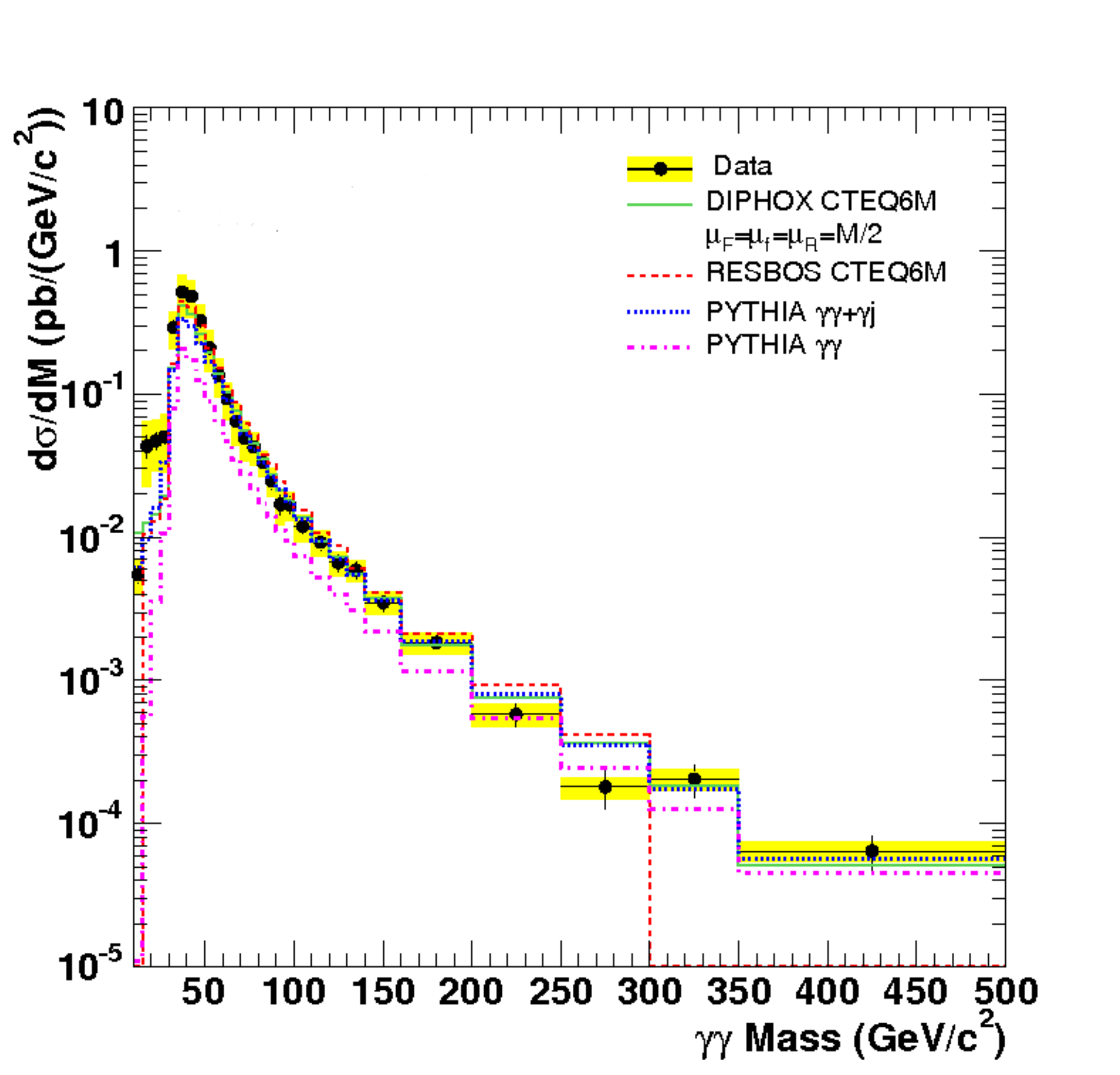}
\caption{CDF measurement of the diphoton invariant mass dependence 
of the diphoton cross section.
\label{fig:cdfdiphoton1}}
\end{figure}

Earlier in Run II, the CDF and D\O\ experiments had each measured 
the inclusive isolated photon cross 
section~\cite{Aaltonen:2009ty,Abazov:2005wc}.
The CDF results are shown in Fig.~\ref{fig:cdfinclphoton1}
as a function of the photon transverse energy.
The pQCD predictions are computed in next-to-leading order (NLO) 
in $\as$ using the program {\sc jetphox}~\cite{Binoth:1999qq,Catani:2002ny}
with CTEQ6.1M PDFs~\cite{Pumplin:2002vw}, 
and renormalization and factorization scales set to 
$\mu_{R,F}=E_T^\gamma$.
The $E_T^\gamma$ dependence of the data/theory ratio is consistent
with what was observed by D\O, and by previous isolated photon measurements
in collider and fixed target experiments at lower energies.

To investigate the potential sources of disagreement
in more detail, the D\O\ collaboration has subsequently
measured this cross section more differentially~\cite{Abazov:2008er}.
For this purpose, also the jet (produced in association with 
the photon) was tagged, and the photon plus jet cross section 
was measured in four regions of the photon and jet rapidities:
For central jets ($|y_{\rm jet}| < 0.8$) 
and for forward jets ($1.5 < |y_{\rm jet}| < 2.5$), and in each case
for same side photons ($y_\gamma \cdot y_{\rm jet} >0$)
and for opposite side photons ($y_\gamma \cdot y_{\rm jet} <0$).
Again, in none of these rapidity regions
could theory describe the $E_T^\gamma$ dependence observed in data.

In an additional measurement~\cite{Abazov:2009de}, the D\O\ collaboration
has measured the photon plus jet production cross section,
this time, however, for heavy flavor jets
(separately for $c$- and $b$-jets, both at central rapidities).
While the photon plus $c$-jet cross section is not described by theory,
NLO pQCD theory is in good agreement with the measured 
photon plus $b$-jet cross sections.
This result is consistent with a measurement of the photon plus $b$-jet 
cross section by the CDF collaboration~\cite{Aaltonen:2009wc}.

A possible conclusion could be that the additional hard scale,
provided by the heavy quark, 
improves the fixed-order pQCD theory calculation.

\begin{figure}
\centering%
\includegraphics[scale=0.4]{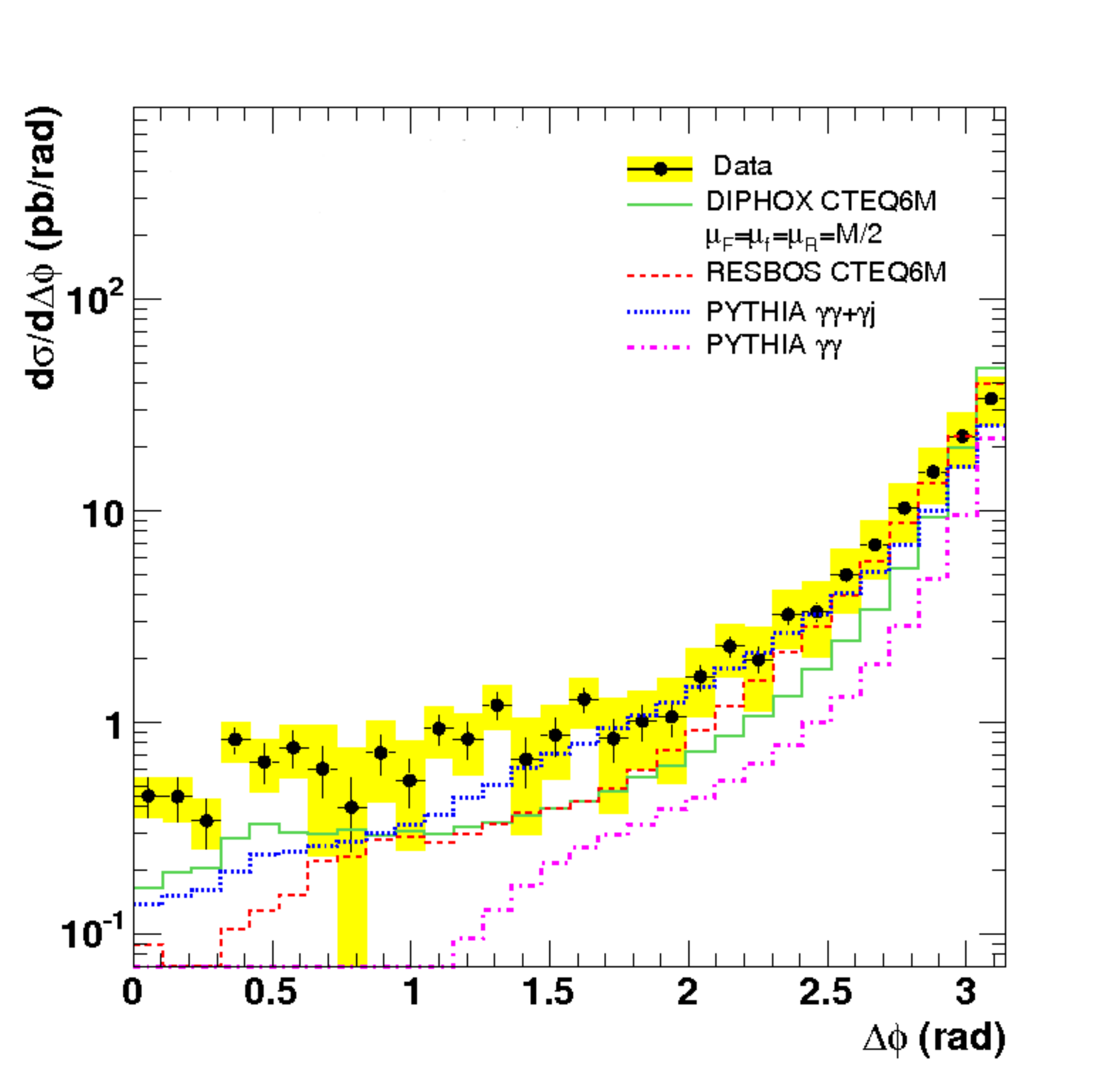}
\caption{CDF measurement of the diphoton $\Delta \phi$ distribution 
of the diphoton cross section.
\label{fig:cdfdiphoton2}}
\end{figure}

\subsection{Diphoton Cross Section}

The leading contributions to diphoton production are from 
quark-antiquark annihilation and from gluon-gluon-scattering.
Although the latter process is suppressed by a factor of $\alpha_s^2$ 
(as the photons couple to a quark box), its contribution is still 
significant at small diphoton masses.
In this kinematic range the PDFs are probed at small momentum fractions,
where the gluon density is much larger than the quark densities.
In addition to the prompt contributions, the diphoton cross section 
also receives contributions where one or two photons are produced 
in fragmentation processes.

The CDF and D\O\ collaborations have both measured 
the diphoton cross section~\cite{Aaltonen:2011via,Abazov:2010ah}
as a function of diphoton invariant mass, diphoton transverse momentum,
and the azimuthal angle difference $\Delta \phi$ between the two photons.
The results are compared to different approximations of pQCD.
The program {\sc diphox}~\cite{Binoth:1999qq} includes NLO matrix elements 
for both the direct contribution and the fragmentation contribution.
The contributions from fragmentation processes are especially large
in the regions of small diphoton masses, large $p_T$, and small $\Delta \phi$.
The program {\sc resbos}~\cite{Balazs:1997hv} has implemented 
the fragmentation contributions only at LO, but it includes 
resummation of soft initial-state gluon radiation
which are especially relevant
at low diphoton $p_T$ and at large $\Delta \phi$. 
The data are also compared to the results from \pythia~\cite{pythia} 
(LO matrix elements plus parton shower and fragmentation model).
The diphoton invariant mass and $\Delta \phi$ distributions are shown in
Figs.~\ref{fig:cdfdiphoton1} and~\ref{fig:cdfdiphoton2}, respectively.
The {\sc diphox} (solid line) and {\sc resbos} (dashed line) predictions both 
give a reasonable overall description of the data, 
except in specific ``critical kinematic regions'' in which
the unique features of the different calculations are probed.
The diphoton mass distribution is described by both
for masses above the peak at $30\,$GeV, while 
both underestimate the cross section at lower masses.
For the diphoton $p_T$ distribution (not shown) only {\sc resbos}
can describe the data for $p_T < 20\,$GeV, the region where 
soft gluon resummation is most important.
Discrepancies between data and theory are most prominent in the comparison
of the measured and predicted distributions of $\Delta\phi$,
where none of the predictions can describe the whole spectrum.
However, in the region $\Delta \phi \rightarrow \pi$,
where soft gluon processes are expected to become relevant,
the {\sc resbos} prediction agrees better with the data.
At smallest values, $\Delta \phi < 1$, corresponding to the region
of smallest diphoton masses, the {\sc diphox} prediction
which includes the additional fragmentation contributions
is closer to the data.

For a better overall description of diphoton production,
it would be desirable to have a single calculation which includes
all of the existing pieces, the full NLO calculations of the 
direct and fragmentation contributions, combined with 
${\cal O}(\alpha_s^3)$ $gg\rightarrow \gamma \gamma$
corrections plus resummed initial-state contributions.

\begin{figure}
\centering%
\includegraphics[scale=0.36]{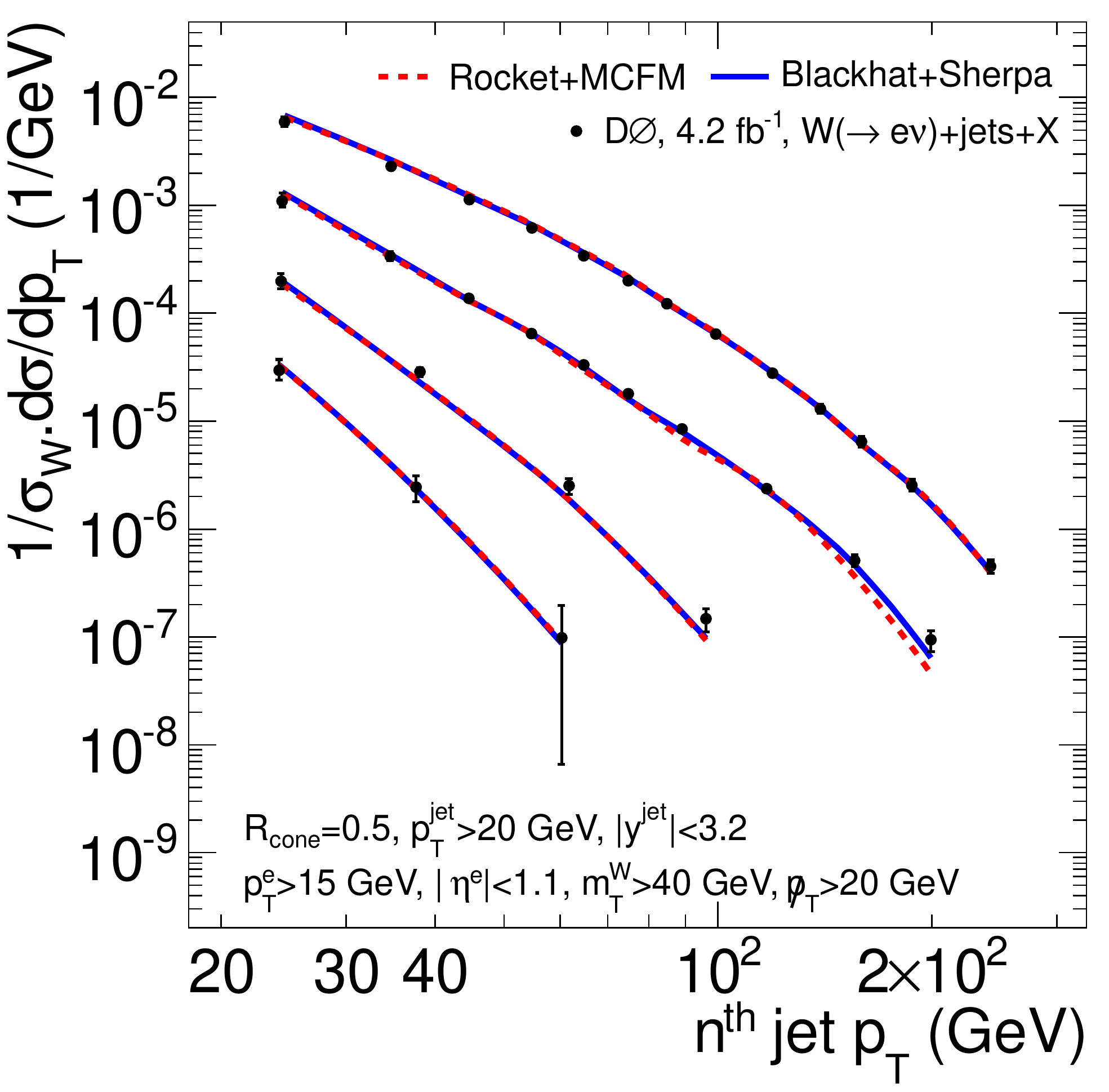}
\caption{D\O\ measurement of the production cross section for
 $W$ + $n$ jets, differentially in $p_T$ of the jets.
The data are compared to different pQCD calculations
in NLO (for $n \le 3$) and LO (for $n=4$).
\label{fig:wjets1}}
\end{figure}

\section{Vector Boson plus Jet Production}
\label{sec:vplusjet}

\begin{figure*}
\centering%
\includegraphics[scale=0.41]{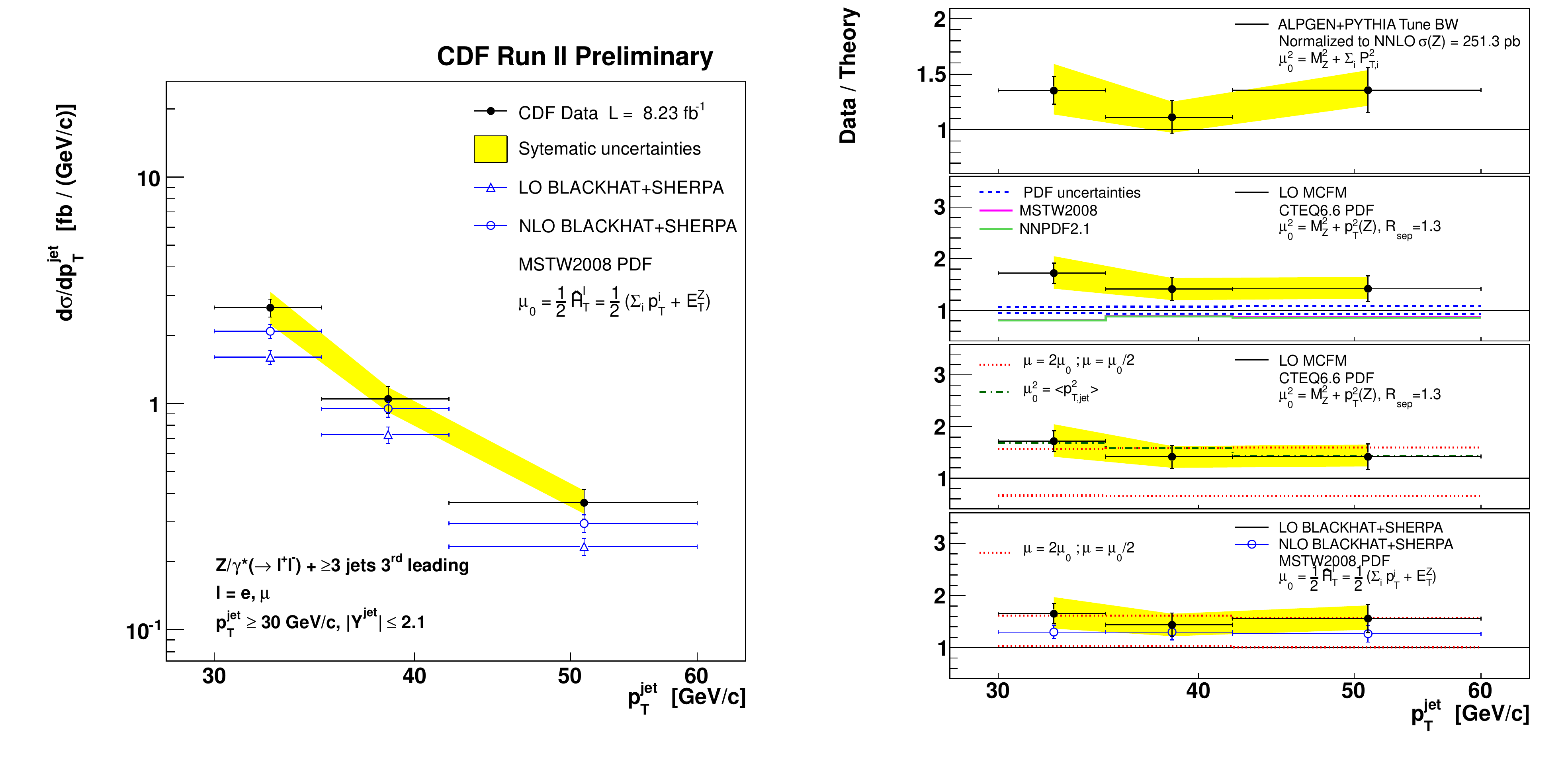}
\caption{CDF Measurement results for the production cross section for
 $Z/\gamma^*$ + 3 jets, differentially in $p_T$ of the third jet (left).
The data are compared to pQCD predictions in LO and NLO.
Ratios of data and theory are shown on the right.
The ratios to LO and NLO pQCD are in the bottom figure 
(``{\sc blackhat+sherpa}''). 
\label{fig:zjets3}}
\end{figure*}

Measurements of the production rates of vector bosons plus jets
provide fundamental tests of pQCD in different approximations.
Available predictions include calculations in fixed order in $\as$,
MC generators using leading order matrix elements plus parton shower, 
or matched tree-level matrix elements combined with parton showers.
While the former are fundamental for precise quantitative tests
of the theory, the latter are important tools for estimating backgrounds
in many searches of new physics signals.
These processes can also constitute the dominant backgrounds for 
single top and $t\bar{t}$ production, for searches of the
Higgs boson, and for searches of new physics signatures beyond 
the standard model.
Especially, $Z$ boson plus $b$ jet production is a significant background
for associated Standard Model Higgs production 
($Z H \rightarrow Z b \bar{b}$) and for searches of 
supersymmetric partners of the $b$ quark.

The production rates of heavy vector bosons accompanied by jets 
have been studied in Run II in a large number of measurements.
These include measurements of jet $p_T$ spectra and angular distributions
of the jet or the jet-$Z/W$ 
system~\cite{Aaltonen:2007ip,:2007cp,Abazov:2009av,Abazov:2009pp}.
Further analyses have measured the production rates of a $Z/W$
accompanied by a heavy ($c$ or $b$ quark) 
jet~\cite{Aaltonen:2008mt,Aaltonen:2009qi,Abazov:2010ix,Abazov:2008qz}.
Here, two recent measurements of $Z$ + jet and $W$ + jet production
by the CDF and the D\O\ collaboration, respectively,
are presented.
Both collaborations have measured jet $p_T$ spectra in 
$Z/W$ + $n$ jet production for $n=1,2,3, 4$.
These measurements constitute the first tests of NLO pQCD predictions
for $Z/W$ + 3 jet production.

\subsection{$W$ plus Jet Production}

The inclusive $W$ $(\rightarrow e\nu_e)$ plus $n$ jet production cross section
is measured in the D\O\ experiment for $n=1,2,3,4$
using a data sample corresponding to 4.2\,fb$^{-1}$~\cite{Abazov:2011rf}.
Jets are reconstructed using the Run II midpoint cone jet algorithm
with cone radius $R=0.5$.
The $W$ + $n$ jet events are selected by requirements on 
the central electron ($p_T > 15\,$GeV), 
the missing transverse energy ($\met > 20\,$GeV),
the transverse mass of the $W$ boson candidate ($M_T^W > 40\,$GeV),
and the jet transverse momenta ($p_T > 20\,$GeV).
In addition, the spatial distance (in $\eta$, $\phi$) 
between the electron and the nearest jet is required to be 
$\Delta R > 0.5$.
Acceptance corrections and background contributions 
from $Z$ + jets, $t\bar{t}$, diboson, 
and single top production are estimated using different 
Monte Carlo event generators~\cite{Abazov:2011rf}.
The backgrounds from multijet production are determined using 
a data driven method.
After background subtraction, the differential distributions in 
jet $p_T$ are corrected for experimental effects,
and the corrected differential cross sections are normalized to the
measured inclusive $W$ boson cross section.
The results are displayed in Fig.~\ref{fig:wjets1}
as a function of $p_T$
(separately for the first, second, third and fourth jet)
and compared to pQCD predictions in NLO (for $n \le 3$)
or LO (for $n=4$) which have been corrected for non-perturbative effects.
The theory predictions have been obtained using the generators
{\sc blackhat+sherpa}~\cite{Berger:2009ba} (solid line) and 
{\sc rocket + mcfm}~\cite{Ellis:2008qc,Giele:2008bc} (dashed line).
In general, within their uncertainties, the NLO pQCD predictions 
reproduce the data, except in a few regions 
(the low $p_T$ region for $W$ + 1 jet  where theory is slightly higher,
and the high $p_T$ region for $W$ + 3 jets where theory is slightly 
lower than the data).
The LO predictions for $W$ + 4 jets describe the data well,
but they are subject to huge uncertainties from the 
renormalization scale dependence.

\begin{figure}
\centering%
\includegraphics[scale=0.40]{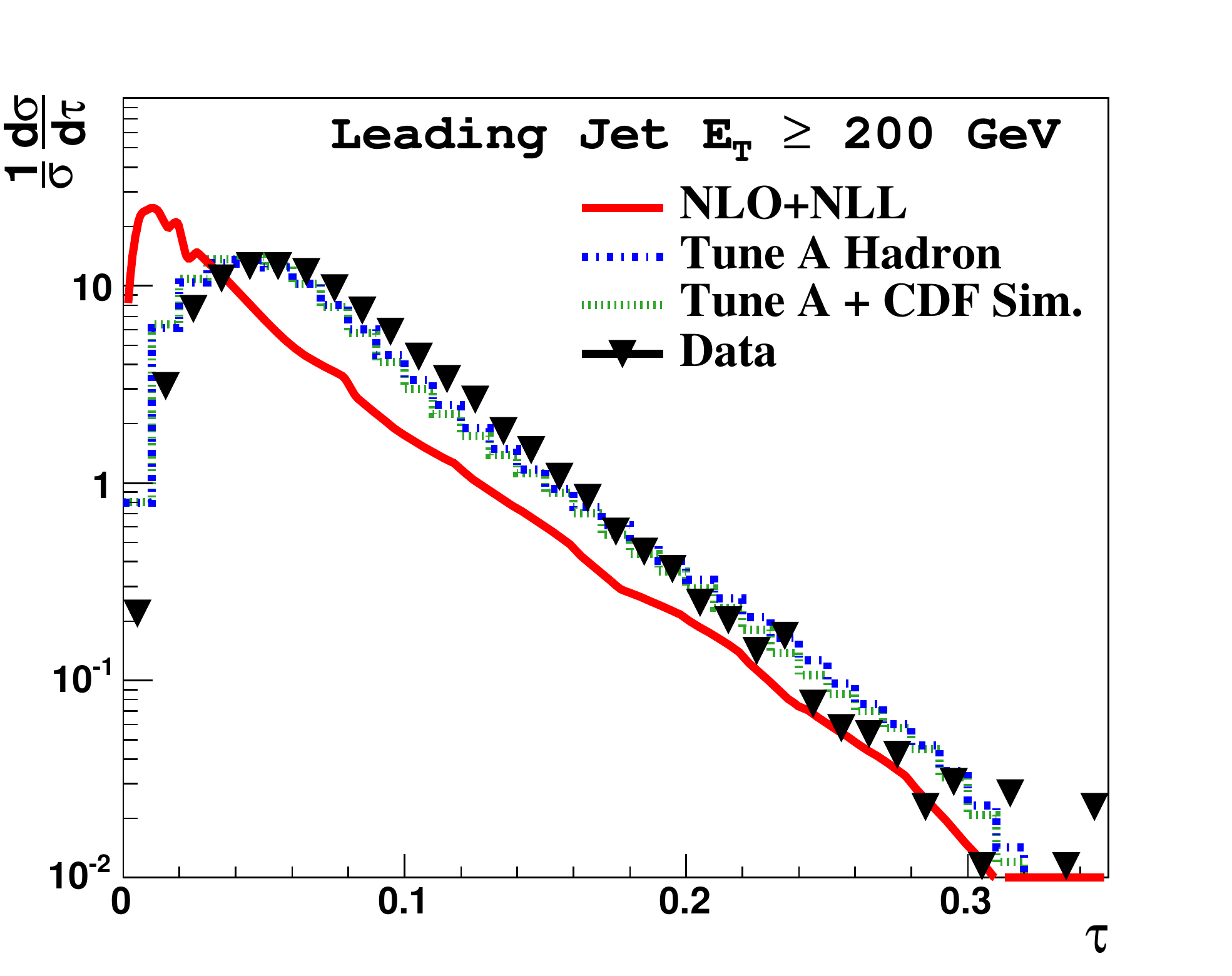}
\caption{CDF measurement of the differential distribution 
of the event shape variable transverse thrust.
\label{fig:evtshp1}}
\end{figure}

\subsection{$Z$ plus Jet Production}

The CDF collaboration has presented a preliminary measurement
of the $Z/\gamma^*$ ($\rightarrow l^+l^-$) + $n$ jet cross section 
(for $n$=1,2,3,4), where $l^\pm = e^\pm$ or $\mu^\pm$~\cite{cdfzjetprel2011}.
A comprehensive set of differential cross sections is measured, 
including distributions of the jet multiplicities,
the jet rapidities, the jet transverse momenta, and
the sum of all jet transverse momenta.
Furthermore, the $Z/\gamma^*$ + 2 jet cross section
has been measured differentially in a set of variables, related to the
dijet system ($M_{jj}$, $\Delta R_{jj}$, $\Delta \phi_{jj}$, $p_T^{jj}$).  
The analysis is based on data corresponding to an integrated 
luminosity of 8.2 fb$^{-1}$.
Jets are reconstructed using the Run II midpoint algorithm with
cone size $R=0.7$, for $p_T > 30\,$GeV and $|y|<2.1$.
The leptons from the $Z/\gamma^*$ decay are required to 
have $p_T> 20\,$GeV and $|\eta|<1.0$ and 
a spatial distance to the nearest jet of $\Delta R > 0.7$.
The measurements for the electron and muon channels are performed 
independently and are combined taking into account asymmetric uncertainties.
The data are corrected for detector effects and are compared to
predictions from pQCD at NLO (for $n$ = 1,2,3) 
or LO ($n=4$) from {\sc blackhat+sherpa} which are corrected for 
non-perturbative effects.
While the $p_T$ and rapidity distributions for 
$Z/\gamma^*$ + 1 jet and for $Z/\gamma^*$ + 2 jet
are reasonably well described by theory,
theory predicts a significantly lower cross section for
$Z/\gamma^*$ + 3 jet production.
The $p_T$ distribution for the third jet is shown in Fig.~\ref{fig:zjets3},
compared to the theory results in LO and NLO.
The large $k$ factor (defined as the ratio of NLO and LO) suggests
a poor convergence of the perturbative expansion, and that
the failure of NLO could easily be due to missing higher order terms.

\section{Event Shapes}
\label{sec:eventshapes}

\begin{figure}
\centering%
\includegraphics[scale=0.40]{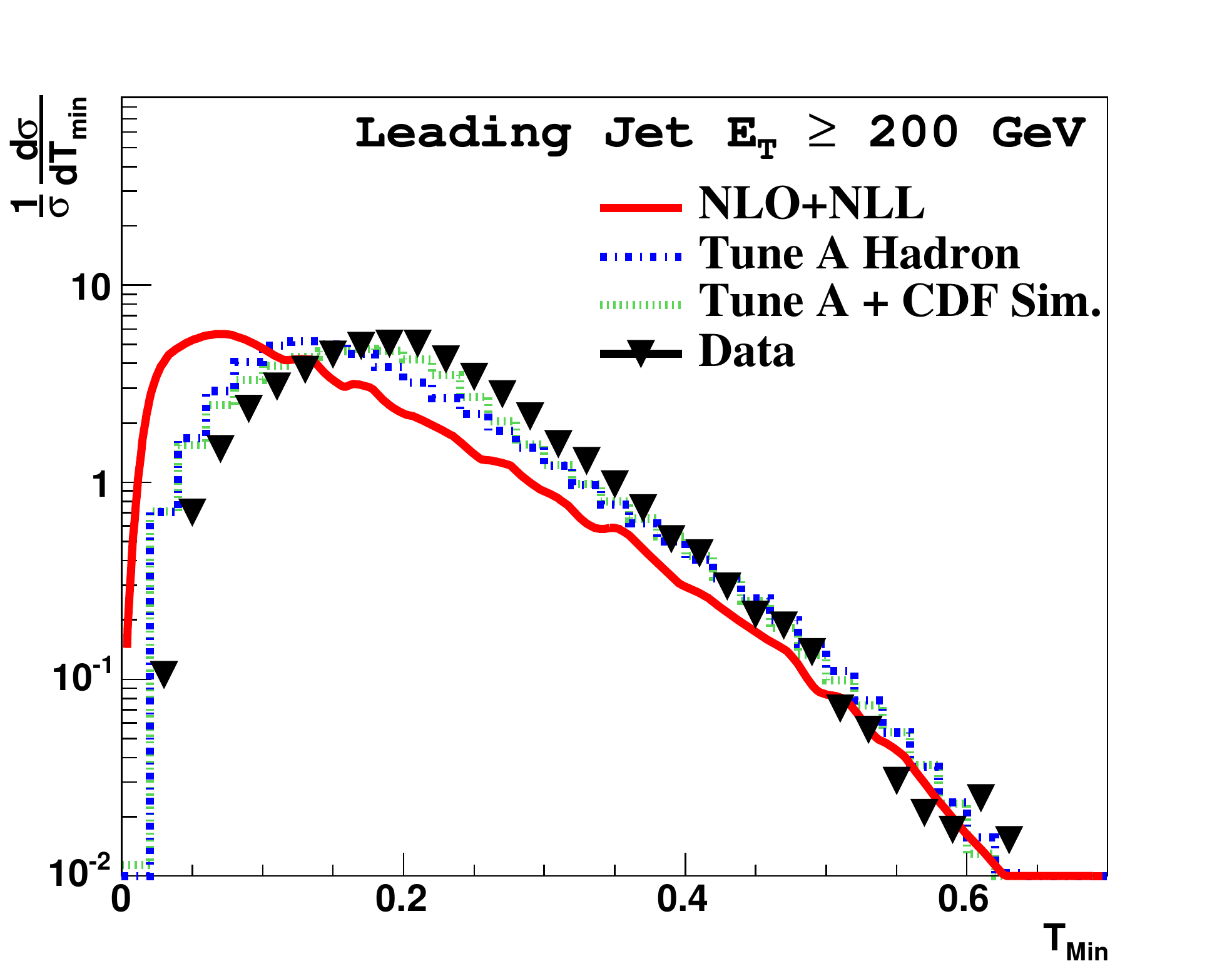}
\caption{CDF measurement of the differential distribution 
of the event shape variable transverse thrust minor.
\label{fig:evtshp2}}
\end{figure}

Event shape variables have been successfully used in $e^+e^-$ collisions
for studying properties of the strong interaction 
and for determinations of $\as$.
Recently, resummed theory predictions (NLO+NLL) for event shape variable in
hadron-hadron collisions have become available~\cite{Banfi:2004nk}.
The CDF collaboration has published measurements of 
the variables thrust and thrust minor in a data set corresponding
to 0.385\,fb$^{-1}$~\cite{Aaltonen:2011et}.
Jets are defined by an iterative cone jet algorithm (without midpoints).
The event shape analysis is performed in an inclusive dijet event sample 
with both leading jets in the central region ($\eta < 0.7$)
and with $E_T^{\rm max} > 200\,$GeV.
The event shape variables are defined in the transverse plane,
over all final state momenta in the full detector acceptance
of $|\eta|<3.5$.
Differential distributions of transverse thrust $\tau$
and transverse thrust minor $T_{\rm min}$
are shown in Figs.~\ref{fig:evtshp1} and~\ref{fig:evtshp2}.
The detector-level data (markers) are compared to \pythia~\cite{pythia}
(with tune A) on particle level (``Tune A Hadron'') 
and with a detector simulation applied (``Tune A + CDF Sim.'').
The small difference between the two model predictions 
indicates that detector effects for these event shapes are small.
It is seen that while \pythia\ with tune A describes the rough features 
of the distributions, it does not describe the tails towards small 
and large values very well.
A parton-level theory prediction in NLO+NLL (solid line),
computed using \nlojet~\cite{Nagy:2003tz,Nagy:2001fj}
and {\sc caesar}~\cite{Banfi:2004yd,Banfi:2003je},
is also overlaid on the data.
This calculation does neither include hadronization nor underlying
event effects.
It was shown in the CDF publication~\cite{Aaltonen:2011et}
that hadronization corrections are actually small for these variables,
but underlying event effects are large.
Therefore one can not expect the NLO+NLL calculation to describe the data.

\begin{figure}
\centering%
\includegraphics[scale=0.40]{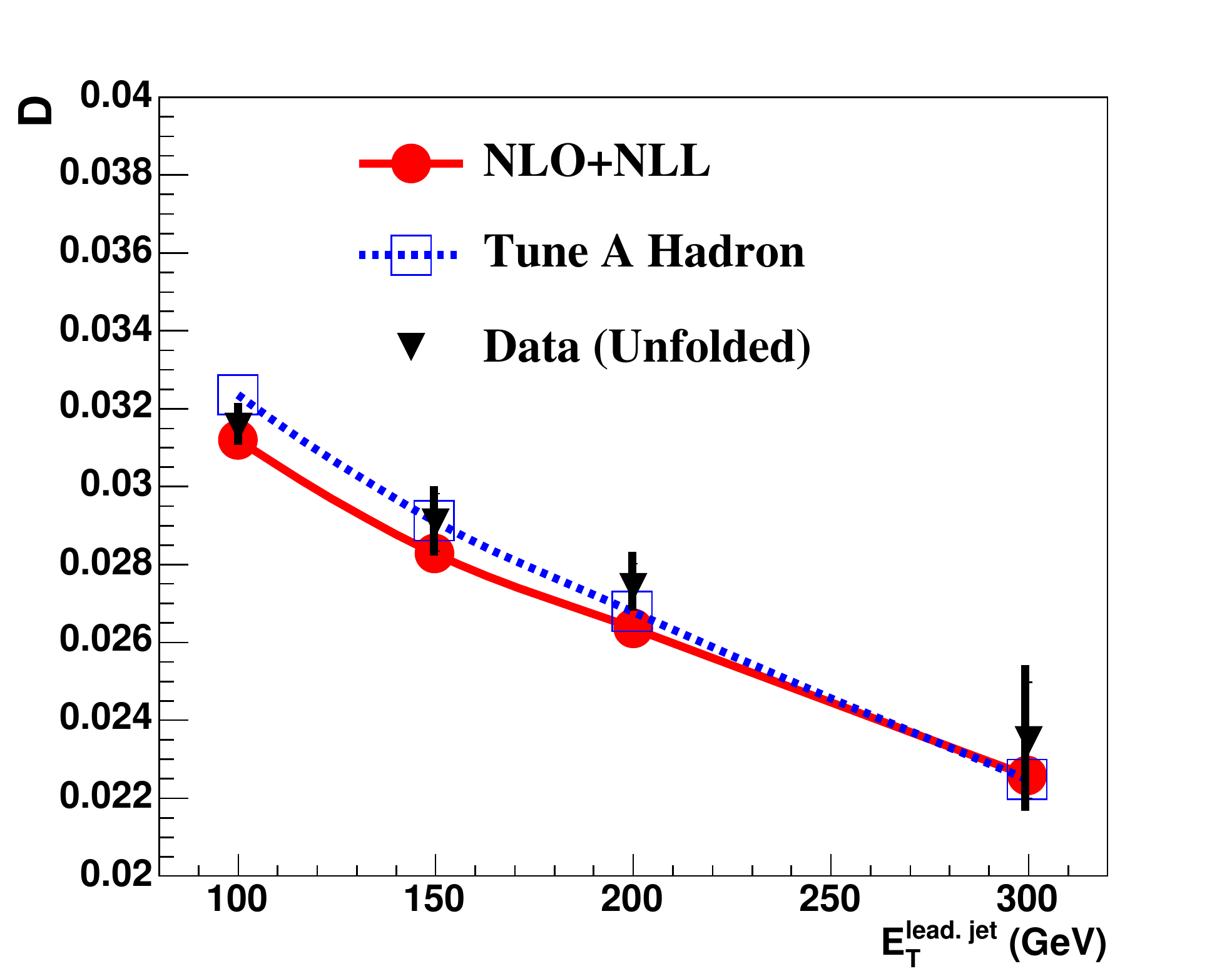}
\caption{CDF measurement of the ``thrust differential'' distribution
(see text)
as a function of the transverse energy of the leading jet.
\label{fig:evtshpD}}
\end{figure}

The CDF collaboration has also proposed and measured a new observable,
defined as the weighted difference of the mean values of 
thrust and thrust minor.
From analyzing the contributions from hard and soft processes
(the latter are assumed to stem from the underlying event)
to the mean values of thrust and thrust minor,
the authors have identified weighting factors $\alpha$ and $\beta$
for the averages of both event shapes, such that their weighted difference 
$( \alpha \langle \tau \rangle - \beta \langle T_{\rm min} \rangle )$
is less sensitive to the soft contributions from the underlying event.
An additional correction factor $\gamma_{\rm MC}$ for underlying
event contributions was computed using \pythia\ tune A
and the variable ``thrust differential'', $D$, was then defined as
$D = \gamma_{\rm MC} 
(\alpha \langle \tau \rangle - \beta \langle T_{\rm min} \rangle)$.
This variable $D$ was measured as a function of $E_T^{\rm max}$,
the leading jet transverse energy in the event.
The results, corrected to particle level, 
are displayed in Fig.~\ref{fig:evtshpD} (markers)
and compared to the particle level prediction from \pythia\ tune A
(dotted line) and to the perturbative (parton-level) prediction
computed in NLO+NLL (solid line).
Both predictions give a good description of the data within
the theoretical uncertainties of $20\%$.

\begin{figure*}
\centering%
\includegraphics[scale=0.92]{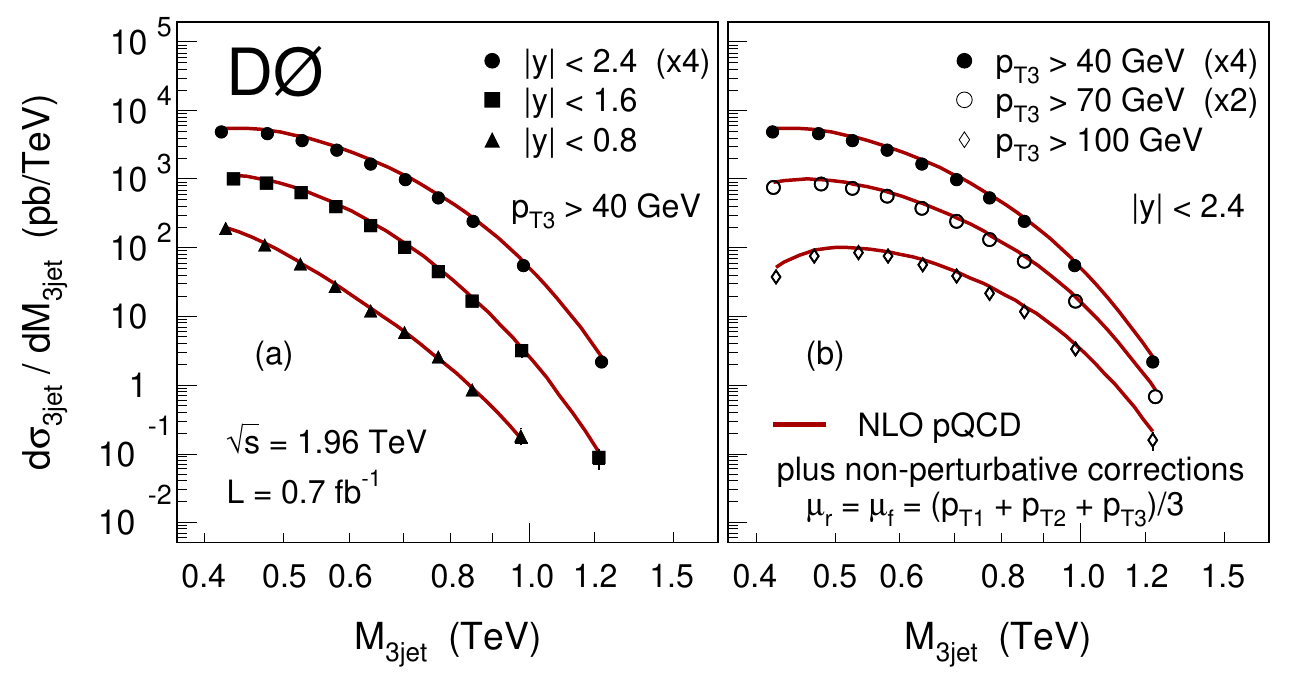}
\caption{D\O\ measurement of the three-jet cross section as a 
function of the three-jet invariant mass in different rapidity regions (left),
and for different $p_{T3}$ requirements (right).
\label{fig:m3j1}} 
\end{figure*}

\section{Multijet Production}
\label{sec:jet}

In Run II of the Tevatron, a large number of fundamental jet observables 
has been measured so far.
The inclusive jet cross section has been measured by the 
CDF and the D\O\ experiments~\cite{Aaltonen:2008eq,:2008hua,Abazov:2011vi}, 
both using the Run II cone jet algorithm with cone size $R=0.7$,
while CDF has also presented results for the 
$k_T$ algorithm~\cite{Abulencia:2007ez} for a distance parameter of 
$R=0.7$, but also for $0.5$ and $1.0$.
The results are presented as double differential cross sections
in jet transverse momentum $p_T$ and rapidity $y$
up to $p_T > 500\,$GeV.
The dijet cross section has been measured by CDF and D\O\ 
as a function of dijet invariant mass~\cite{Aaltonen:2008dn,Abazov:2010fr} 
(the D\O\ result was also measured
as a function of dijet rapidity $|y_{\rm max}|$).
Dijet angular distributions have been measured by the 
D\O\ experiment
in the angular variable $\chijj = \exp(|y_1-y_2|)$
and in different regions of dijet invariant mass, 
up to above 1\,TeV~\cite{:2009mh}.

The analysis of the measured dijet angular distributions has shown 
no indications for new physics processes which would modify the 
angular distributions of jets at high dijet invariant masses, 
such as quark compositeness and extra spatial dimensions.
This analysis has produced the best pre-LHC limits on quark 
compositeness, ADD Large Extra Dimensions, 
and TeV$^{-1}$ Extra Dimension models.
Furthermore, the dijet mass spectra show no evidence for 
resonances, produced by new heavy particles, decaying into jets.
The CDF collaboration has therefore used the dijet mass spectrum 
to set stringent limits on hypothetical particles
such as excited quarks, axigluons, flavor-universal colorons, 
E6 diquarks, color-octet technirhos, or $W'$ and $Z'$ bosons. 
Being confident that no new physics processes contribute
to the measured jets, the inclusive jet data from both experiments
have been used in global 
PDF analyses~\cite{Martin:2009iq,Lai:2010vv,Ball:2011mu}
exploiting the unique sensitivity of the jet data to the gluon density
in the proton at high values of the proton momentum fraction $x$.
Later, however, it has been seen that the rapidity dependence of the
D\O\ dijet data is not well described by all PDFs which 
were obtained from PDF analyses including the inclusive
jet data~\cite{Abazov:2010fr}.

Based on the existing studies, one can conclude that 
inclusive jet and dijet production processes are well understood
and modeled adequately in pQCD.
It is therefore attractive to extend the QCD jet studies towards 
larger jet multiplicities probing higher orders in pQCD.
Early in Run II, the D\O\ collaboration has measured 
dijet azimuthal decorrelations~\cite{Abazov:2004hm}, which was a first tests
of pQCD predictions using the three-jet matrix elements 
which have been computed to NLO~\cite{Nagy:2003tz,Nagy:2001fj}.
Recent measurements extend these studies to different three-jet observables.
The D\O\ collaboration has published a measurement of the three-jet cross
section~\cite{Abazov:2011ub}, multi-differentially 
in the three-jet invariant mass,
$M_{3j}$, in the rapidity region for the three jets, 
and in the $p_T$ requirement for the third jet $p_{T3}$.
A preliminary D\O\ result measures the ratio of three-jet and dijet
cross sections as a function of the leading jet $p_T$ and for different
$\ptmin$ requirements for the other jets.

\subsection{Three-Jet Cross Section}

Jet are defined by the Run II midpoint cone algorithm with cone size
$R=0.7$.
The rapidities of the three leading $p_T$ jets are restricted to
$|y| < 0.8$, $|y| < 1.6$, or $|y| < 2.4$, alternatively.
The $p_T$ requirements are $p_{T1} > 150$\,GeV and $p_{T3} > 40$\,GeV
(with no further requirement for $p_{T2}$).
Additional measurements are made for $p_{T3} > 70$\,GeV 
and $p_{T3} > 100$\,GeV (both for $|y| < 2.4$).
In all cases, all pairs of the three leading $p_T$ jets are required 
to be separated by 
$\Delta R = \sqrt{(\Delta y)^2 + (\Delta \phi)^2}>  2 \cdot \Rcone = 1.4$.
This separation requirement reduces the phase space 
in which pairs of the three leading $p_T$ jets are subject to 
the overlap treatment in the cone jet algorithm.
Since this overlap treatment can strongly depend on details
of the energy distributions in the overlap area,
this region of phase space may not be well modeled by pQCD
calculations at lower orders.
In the remaining analysis phase space, NLO pQCD calculations
are not affected by the Run II cone algorithm's 
infrared sensitivity~\cite{Salam:2007xv}.

\begin{figure*}
\centering%
\includegraphics[scale=0.92]{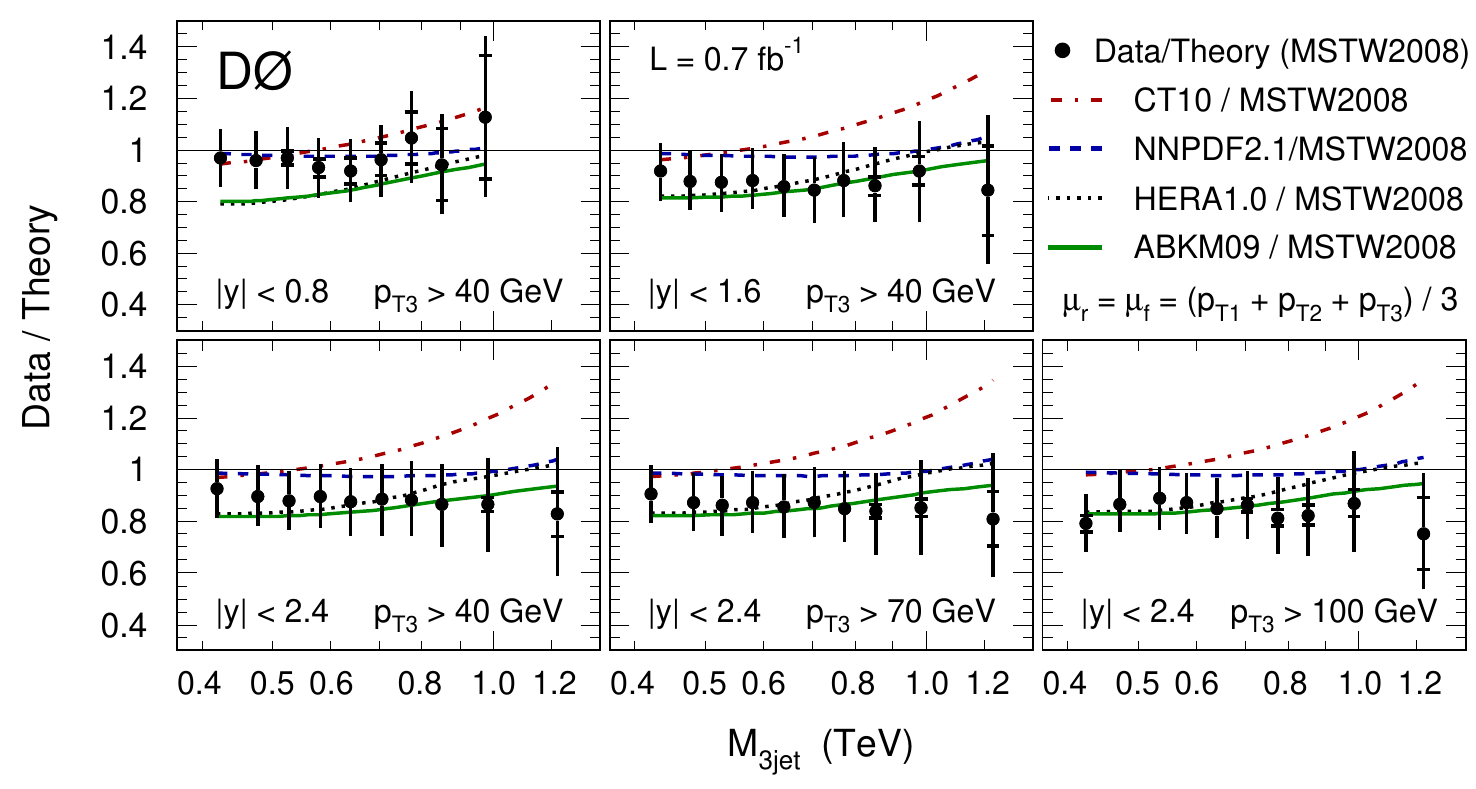}
\caption{Ratios of data and theory for the D\O\ three-jet cross section 
measurement.
The ratios are computed for different PDFs and are shown as 
a function of the three-jet invariant mass 
in different rapidity regions and for different $p_{T3}$ requirements.
\label{fig:m3j3}} 
\end{figure*}

These results~\cite{Abazov:2011ub}, displayed in Fig.~\ref{fig:m3j1}, show the
three-jet cross section, differentially in $M_{3j}$ for different
rapidity requirements (left) and for different $p_{T3}$ requirements (right).
The data are compared to theory predictions which combine
NLO pQCD results (computed using \fastnlo~\cite{Kluge:2006xs}, 
based on \nlojet~\cite{Nagy:2003tz,Nagy:2001fj}) and non-perturbative
corrections (computed using \pythia\ with tune DW~\cite{Albrow:2006rt}).
The comparison is made for MSTW2008NLO PDFs~\cite{Martin:2009bu}
and the corresponding
value of $\asmz = 0.120$.
The renormalization and factorization scales
are set to the average $p_T$ of the three leading $p_T$ jets
$ \mu_R = \mu_F = \mu_0 = (p_{T1}+p_{T2}+p_{T3})/3$.
Ratios of data and theory are shown in Fig.~\ref{fig:m3j3} 
for different PDFs.
For MSTW2008NLO PDFs, the ratios of data and theory are almost constant,
with only a small dependence on $M_{3j}$ and the $|y|$ and $p_{T3}$ 
requirements.
The central data values are below the central theory predictions
by approximately (4--15)\% in the different scenarios,
slightly increasing with $|y|$ and with $p_{T3}$.
In all cases, the data lie inside the range covered by the 
renormalization and factorization scale uncertainties (not shown here).
Theory for CT10 PDFs~\cite{Lai:2010vv}  
(and the corresponding value of $\as(M_Z)=0.118$)
predicts a different shape for the $M_{3j}$ dependence of the cross section.
For $M_{3j} < 0.6\,$TeV, the results for CT10 PDFs
agree with those for MSTW2008NLO, while the CT10 predictions
at $M_{3j} = 1.2\,$TeV are up to 30\% higher.
Further theory results are compared to the data for other PDFs,
including
NNPDFv2.1~\cite{Ball:2011mu} ($\asmz = 0.119$),
ABKM09NLO~\cite{Alekhin:2009ni} ($\asmz = 0.1179$),
and HERAPDFv1.0~\cite{:2009wt} ($\asmz = 0.1176$).
The results for NNPDFv2.1 agree everywhere within $\pm$4\% with those
from MSTW2008NLO.
The cross sections predicted for HERAPDFv1.0 are (15--20)\% below those
for CT10 everywhere and their $M_{3j}$ distributions have a similar shape.
The $M_{3j}$ dependence of the calculations for the ABKM09NLO PDFs
is between the shapes of MSTW2008NLO/NNPDFv2.1 and CT10/HERAPDFv1.0.
At low $M_{3j}$, the predictions for ABKM09NLO agree with those
for HERAPDFv1.0, while at higher $M_{3j}$,
they predict the smallest cross sections of all PDFs under study.

\begin{figure*}
\centering%
\includegraphics[scale=0.9]{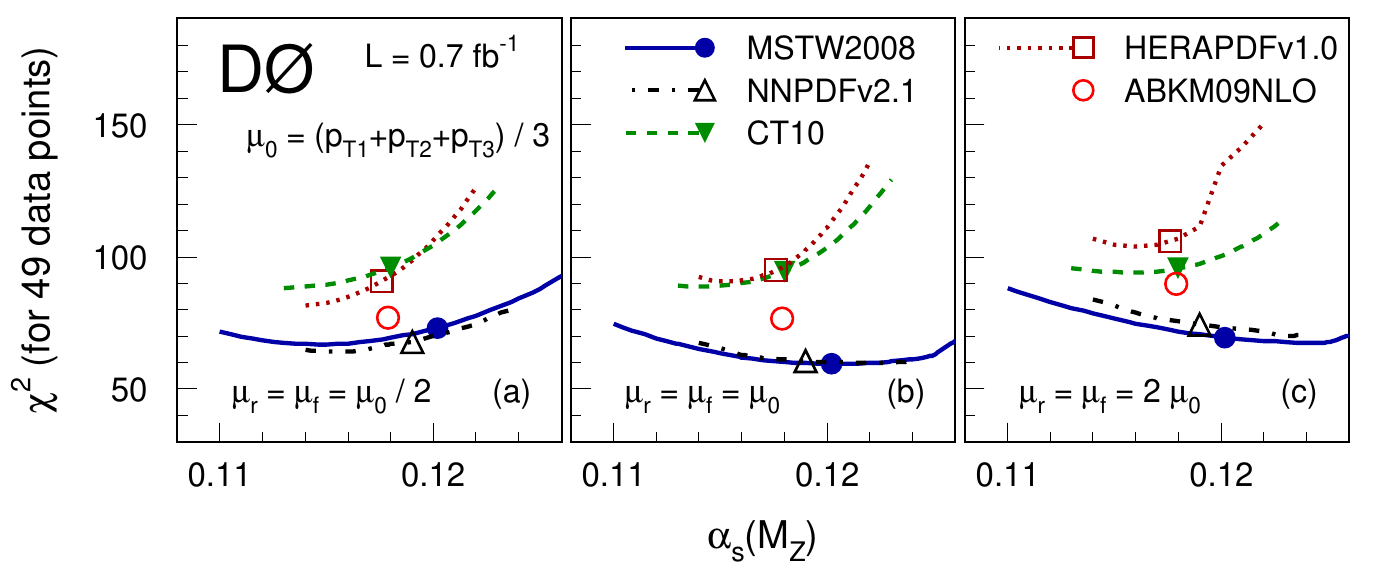}
\caption{The $\chi^2$ values between data and theory 
for the three-jet cross section results, displayed as a function
of $\asmz$ in the theory calculations, for different choices
of the scales (from left to right) and for different PDF
parametrizations (different lines).
\label{fig:m3j4}} 
\end{figure*}

To quantify the agreement between data and theory, a $\chi^2$ 
is computed, fully taking into account the correlations of 
uncertainties~\cite{Abazov:2011ub}.
The $\chi^2$ is computed for different scales (for the nominal 
scale and for variations by factors of 0.5 and 2), for different
PDF parametrization, and for different values of $\asmz$ 
used in the NLO matrix elements and PDFs
(for all $\asmz$ values for which PDF sets are available).
The $\chi^2$ results are shown in Fig.~\ref{fig:m3j4}.
For $\asmz$ values close to the world average of
$0.1184\pm0.0007$~\cite{Bethke:2009jm},
for all PDF sets, with the exception of HERAPDFv1.0,
the lowest $\chi^2$ is obtained for the central choice of the scales.
From all PDFs, the largest $\chi^2$ values are obtained for
CT10 and HERAPDFv1.0 PDFs, independent of the scale and $\asmz$ choices.
The best overall agreement, corresponding to the lowest $\chi^2$ values,
is obtained for MSTW2008NLO
for the central scale choice and $\asmz=0.121$,
and the results for NNPDFv2.1 are very close.

\subsection{Ratio of Three-Jet and Dijet Cross Sections}

\begin{figure*}
\centering%
\includegraphics[scale=1]{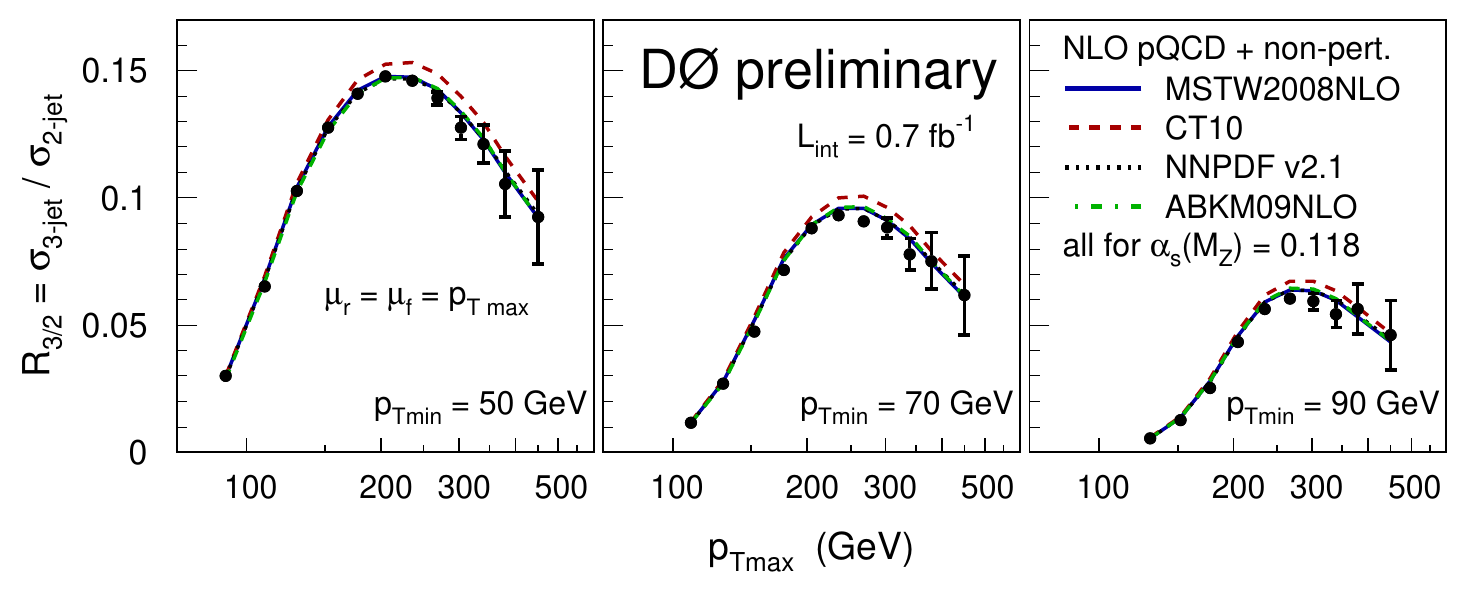}
\caption{Preliminary D\O\ results for the ratio of three-jet and dijet 
cross sections, $\Rtt$, as a function of leading jet transverse momentum 
$\ptmax$ for different $\ptmin$ requirements (from left to right, 
$\ptmin \ge 50, 70, 90\,$GeV).
The data are compared to NLO pQCD predictions for different PDFs.
\label{fig:r32}}
\end{figure*}

As discussed above, the interpretation of the three-jet cross section
(and any other jet cross section) depends strongly on the choices
of $\asmz$ and the PDFs.
The impact of the latter can be strongly reduced, by studying 
observables defined as ratios of multijet cross sections.
The D\O\ collaboration has presented a preliminary result~\cite{D0:2010rr} 
for the ratio of the inclusive three-jet and dijet cross sections, $\Rtt$, 
measured as a function of the leading jet transverse momentum $\ptmax$.
For this choice, in every $\ptmax$ bin, the numerator is
a subset of the denominator, and therefore
the variable $\Rtt$ represents the conditional probability 
that a given inclusive dijet event also has a third jet.
The value of $\ptmax$ is a common scale for the three-jet and 
the dijet production processes.
Therefore $\Rtt(\ptmax)$ is directly sensitive to $\as$ 
at the scale $\mu_R=\ptmax$ while the PDFs cancel to a large extent 
in the cross section ratio.
In this analysis, the $n$-jet cross section (for $n=2,3$) is defined 
by all events with $n$ or more jets with $p_T$ above $\ptmin$,
in a given rapidity region (here: $|y|<2.4$ for the $n$ leading jets),
for $\ptmin = 50,\, 70,\, 90\,$GeV.
The results are displayed in Fig.~\ref{fig:r32} 
as a function of $\ptmax$, for different $\ptmin$ requirements 
(from left to right).
Theory calculations based on NLO pQCD plus non-perturbative corrections
are compared to the data for 
different PDFs (MSTW2008NLO, CT10, NNPDFv2.1, ABKM09NLO)
using in all cases $\asmz=0.118$ (in the matrix elements and in the PDFs).
In all cases good agreement is seen.

\section{Determination of the Strong Coupling Constant}
\label{sec:as}

The strong coupling constant, $\asmz$, is one of the fundamental 
parameters of the Standard Model of Particle Physics.
The energy dependence of $\as$ is predicted by the renormalization group 
equation (RGE).
The value of $\as$ has been determined in many different processes,
including a large number of results from hadronic jet production,
in either $e^+e^-$ annihilation or in deep-inelastic $ep$ scattering 
(DIS) up to energies of $209\,$GeV~\cite{Bethke:2009jm}.
Prior to the analysis presented in this article, however, 
only a single result had been obtained
from jet production in hadron-hadron collisions.
This result is
$\asmz = 0.1178 
        ^{+0.0081}_{-0.0095} (\mbox{exp.}) 
        ^{+0.0071}_{-0.0047} (\mbox{scale}) 
        \pm 0.0059 (\mbox{PDF})    $,
extracted by the CDF collaboration from inclusive jet cross section 
data in $\ppbar$ collisions at $\sqrt{s}=1.8\,$TeV~\cite{Affolder:2001hn}.
All individual uncertainty contributions for this result are larger
than those from comparable results from $e^+e^-$ annihilation or 
DIS~\cite{Bethke:2009jm}.

\begin{figure*}
\centering%
\includegraphics[scale=0.9]{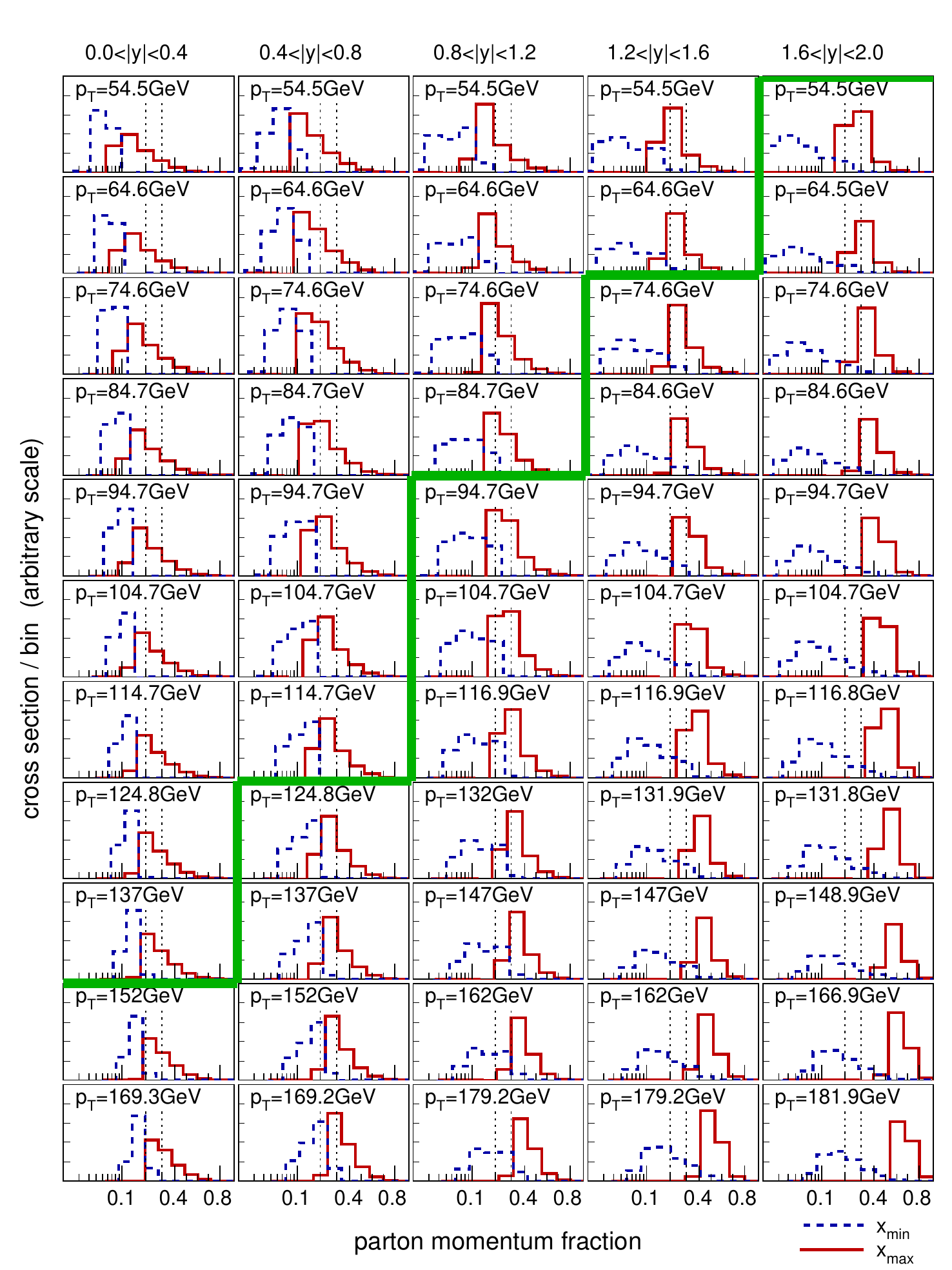}
\caption{Distributions of the (anti-) proton momentum fractions
$x_{\rm max}$ and $x_{\rm min}$ carried  by the initial-state partons,
for a set of $p_T$ and $|y|$ bins of the inclusive jet cross section,
as predicted by NLO pQCD.
The ($p_T$, $y$) bins  which are used in the $\as$ analysis are
the ones to the left (or above) the bold line.
\label{fig:xmax}}
\end{figure*}

Recently, the D\O\ collaboration has presented a new 
$\as$ determination~\cite{Abazov:2009nc}
with unprecedented precision at a hadron collider.
The $\as$ result is extracted from inclusive jet cross section data
in $\ppbar$ collisions at $\sqrt{s}=1.96\,$TeV from 
a recent D\O\ measurement~\cite{:2008hua,Abazov:2011vi}.
The pQCD prediction for the inclusive jet cross section is given by
\begin{equation}
\sigma_{\mbox{pert}}(\as) = 
     \left(  \sum_n \as^n c_n \right) \otimes f_1(\as) \otimes f_2(\as)
     \, ,
\label{eq:pQCD}
\end{equation}
where the $c_n$ are the perturbative coefficients,
the $f_{1,2}$ are the PDFs of the initial state hadrons,
and the ``$\otimes$'' sign denotes the convolution over 
the momentum fractions $x_1$, $x_2$ of the hadrons.
The sum runs over all powers $n$ of $\as$ which contribute to the 
calculation.
The D\O\ result is based on NLO pQCD ($n=2,3$) plus 
2-loop contributions from threshold corrections~\cite{Kidonakis:2000gi}
($n=4$). 
The latter reduce the scale dependence of the calculations,
leading to a significant reduction of the corresponding uncertainties.
While the $f_{1,2}$ have no explicit $\as$ dependence, their knowledge
depends on $\as$ (due to $\as$ assumptions in the PDF analyses).
Since the RGE uniquely relates the value of $\asmur$ at any scale $\mu_R$
to the value of $\asmz$, all equations can be expressed in terms of $\asmz$. 
The total theory prediction for inclusive jet production is given by the pQCD 
result in (\ref{eq:pQCD}), multiplied by a correction factor for 
non-perturbative effects
\begin{equation}
 \sigma_{\mbox{theory}}(\asmz) =  
   \sigma_{\mbox{pert}}(\asmz)
    \cdot  c_{\mbox{non-pert}} \, . \, \, \,
\label{eq:allQCD}
\end{equation}
The pQCD results are computed in \fastnlo, which is based on \nlojet\
and on the calculations from Ref.~\cite{Kidonakis:2000gi}.
To determine $\asmz$, recent PDF results are used
and $\asmz$ is varied in $\sigma_{\mbox{pert}}(\asmz)$
(i.e.\ simultaneously in the matrix elements and in the PDFs)
until $\sigma_{\mbox{theory}}(\asmz)$ agrees with the data.
There are, however, two conceptual issues when extracting $\as$ from
cross section data.
\begin{enumerate}

\item
When performing the DGLAP evolution of the PDFs,
all PDF analyses are assuming the validity of the RGE
which has so far been tested only for energies up to 209\,GeV.
Since extracting $\as$ at higher energies means testing 
(and therefore questioning) the RGE,
using these PDFs as input would be inconsistent.

\item
D\O\ jet data have been used in all recent global PDF analyses.
The PDF uncertainties are therefore correlated with the experimental 
uncertainties in those kinematic regions in which 
the D\O\ jet data had strong impact on the PDF results.
As shown in Figs.~51--53 in Ref.~\cite{Martin:2009bu}, 
this is the case for the proton's  gluon density
at $x > 0.2 - 0.3$.
Since the correlations between PDF uncertainties and 
experimental uncertainties are not documented, 
the $\as$ extraction should avoid
using those data points which already
had significant impact on the PDF results.

\end{enumerate}
In light of the second issue, the D\O\ $\as$ extraction
uses only data points which are insensitive to $x > 0.2 - 0.3$.
Figure~\ref{fig:xmax} shows the cross section contributions
as a function of momentum fractions $x$ 
(here separately for $x_{\rm max}$ and $x_{\rm min}$ in an event) 
for the inclusive jet cross section bins of the D\O\ measurement.
The ($p_T$, $y$) bins in the top left above the solid line
have only small contributions from momentum fractions
$x > 0.2 - 0.3$.
Since all of these data points have $p_T$ below 145\,GeV,
the first issue does not become relevant here.
This leaves 22 (out of 110) inclusive jet data points
in the range $50 < p_T < 145\,$GeV for the $\as$ analysis.

\begin{figure}
\includegraphics[scale=0.95]{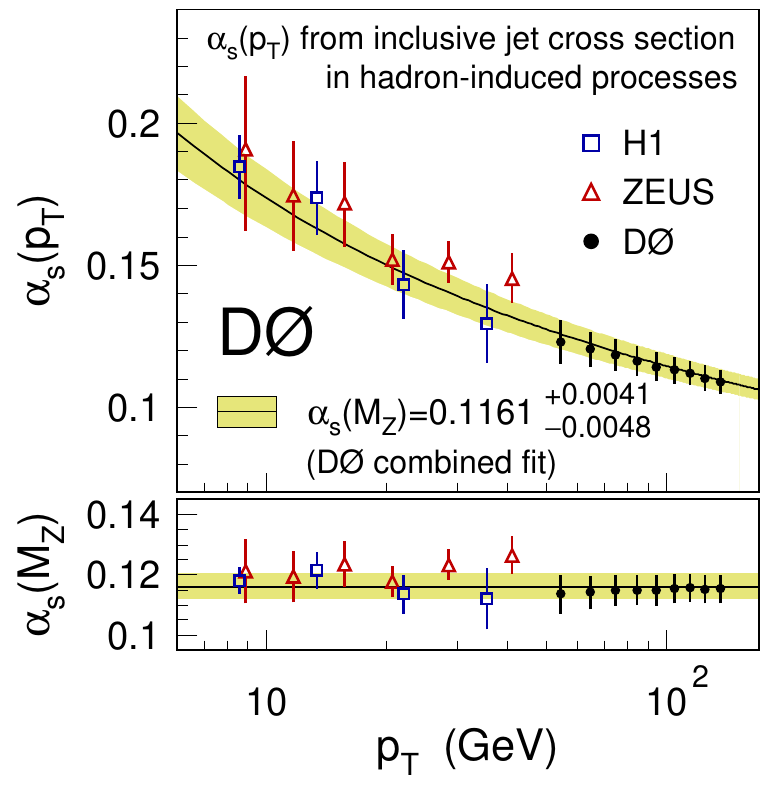}
  \caption{\label{fig:fig1}
   Recent D\O\ results from a determination
   of the strong coupling from inclusive jet cross section data,
   compared to corresponding results in DIS from HERA.
  }
\end{figure}

The $\as$ extraction
uses PDFs from the MSTW2008 analysis~\cite{Martin:2009iq}
which were obtained at NNLO
(consistent with the precision of the 
theory calculation used here).
These PDFs have been determined for 21 $\asmz$ values between
$0.107$ and $0.127$~\cite{Martin:2009bu}.
The continuous $\asmz$ dependence of the pQCD cross sections is
obtained by interpolating the cross section results
for the PDF sets for different $\asmz$ values.
PDF uncertainties are computed using the twenty uncertainty 
eigenvectors (corresponding to 68\%~C.L.).
The uncertainties in the pQCD calculation due to uncalculated higher-order 
contributions are estimated from the $\mu_{R,F}$ dependence of the 
calculations by varying the renormalization and factorization 
scales in the range $0.5 \le (\mu_{R,F}/p_T) \le 2$.
In a first step, data points with the same  $p_T$ are combined to determine
nine values of $\aspT$ for $50 < p_T < 145\,$GeV.
These results are shown in Fig.~\ref{fig:fig1} 
and compared to results obtained in DIS.
A combined determination from all 22 data points yields a result of
\begin{eqnarray*}
\asmz & = & 0.1161  ^{+0.0034}_{-0.0033} \, \mbox{(exp.)}   \\ 
& & \phantom{mmmm} ^{+0.0010}_{-0.0016} \, \mbox{(non-pert.)} \\   
& &  \phantom{mmmm} ^{+0.0011}_{-0.0012} \, \mbox{(PDFs)}   \\
& &  \phantom{mmmm} ^{+0.0025}_{-0.0029} \, \mbox{(scale)}    .
\end{eqnarray*}
This is currently the most precise result from a hadron collider,
with similar precision as recent results from jet production in DIS,
and consistent with the current world average value~\cite{Bethke:2009jm}.

\section{Summary }

The CDF and D\O\ collaborations have measured a large number
of QCD related observables testing many aspects of the standard model
predictions.
The high integrated luminosity from the Tevatron
allows to explore rare processes like 
vector boson plus jet production for the first time in great detail.
While inclusive photon and diphoton measurements are still
challenging theory,
other results are in general well described by existing
theoretical approximations.
Precise measurements of jet observables
have strong impact in precision phenomenology,
as in constraining the proton PDFs, determinations of $\as$,
and for excluding new physics models.








\end{document}